\documentclass[showpacs,aps,pra,epsfigure,twocolumn]{revtex4}
\usepackage{dcolumn}
\usepackage{bm}
\usepackage{graphicx}
\usepackage{amsmath}
\usepackage{latexsym}
\usepackage{amsfonts}
\usepackage{amssymb}
\usepackage{array}
\usepackage{epsfig}
\usepackage{epstopdf}

\newcommand{\one}{\mbox{$1 \hspace{-1.0mm}  {\bf l}$}}

\newcommand{\nbar}{\overline{n}}

\begin{document}

\title{Phase-space behavior and conditional dynamics of an optomechanical system}
\author{S. Gr\"oblacher$^{1}$\footnote{Present address: Institute for Quantum Information and Matter, California Institute of Technology, Pasadena, California, 91125, USA.}, 
S. Gigan$^2$, M. Paternostro$^3$}
\affiliation{$^1$Vienna Center for Quantum Science and Technology, Faculty of Physics, University of Vienna, Boltzmanngasse 5, 1090 Wien, Austria}
\affiliation{$^2$Institut Langevin, ESPCI ParisTech, CNRS UMR 7587, Universit\'e Paris VI \& VII, INSERM, 1 Rue Jussieu, 75005 Paris, France}
\affiliation{$^3$School of Mathematics and Physics, Queen's University, Belfast BT7 1NN, United Kingdom}
\date{\today}

\begin{abstract}
We characterize the local properties of an optomechanical system comprising the movable mirror of a resonator and its intracavity field, mutually coupled via radiation-pressure. Our approach shows that both the state of the mirror and the field can be interpreted as squeezed thermal states whose dynamical properties can be tuned by properly choosing the working parameters. This allows us to design conditional procedures for the amplification of the correlation properties of the optomechanical state. Our study is a step forward in the understanding of the physics that rules a system of current enormous experimental interest.
\end{abstract}
\pacs{03.67.-a,03.67.Dd,03.67.Hk }
\maketitle


The observation of quantum dynamics of a macroscopic system is an outstanding goal in modern physics. 
Several paths have been followed in order to achieve macroscopic quantum mechanical behaviors, using nano-electro-mechanical devices coupled to single-electron transistors or optomechanical systems coupling electromagnetic fields and vibrating mechanical structures~\cite{progressi}. The {\it desideratum} common to all such physical situations is the achievement of negligible thermal excitations of any mechanical system, whose energy has to be pushed down to a level comparable with their specific single-quantum transition energy. Recently, some notable results have been achieved in this respect, both theoretically and experimentally~\cite{david,cool,nano,ionjp,altri}, by exploiting back-action induced mechanisms. Moreover, correlations of quantum mechanical nature have been predicted in micro-optomechanical systems, resilient to the exposure to Markovian environmental noise and relatively high operating temperatures~\cite{david}. Progresses towards the experimental revelation of their classical counterparts have been recently reported~\cite{loro}. This complements previously studied unitary dynamics~\cite{boseferreira} and sets the ground for realistic experimental investigations. Indeed, operative schemes for the observation of optomechanical entanglement have been suggested, based on the use of ancillary fields and all-optical observations~\cite{3Bpaper}. Noticeably, very recently, evidences of quantum effects in the motion of a mechanical system have been observed~\cite{Safavi}.

In the optomechanical setting comprising an optical cavity with a movable mirror coupled to light through radiation pressure forces, however, a complete understanding of the effects that the coupling with light has on a micromechanical device is still lacking. The ``not immediately intuitive'' nature of the correlations studied in~\cite{david,3Bpaper}, for instance, deserves a deeper investigation: the properties of the whole system and its constituents have to be properly unravelled. In this context, in consideration of the formal equivalence between the movable mirror and a quantum harmonic oscillator, it is natural to use techniques and tools of quantum optics to provide a more explicit interpretation of the dynamics of the system. This is exactly the direction considered in this paper, where we draw a phase-space analysis of the cavity mirror-field system depicted above. This allows not only for a complete characterization of both the mirror and the field as squeezed thermal states, but also the identification of conditioning strategies, based on measurement and postselection, through which the state of the mirror is significantly modified. 

 The paper is organized as follows. In Sec.~\ref{formalism} we introduce the formalism and technical details being used in order to tackle the problem at hand. A detailed and extensive description of the methods can be found in Refs.~\cite{david,3Bpaper} and here we only sketch the strategy to follow in order to gather the form of the covariance matrix of each optomechanical system. Sec.~\ref{Wigner} is devoted to the description of the dynamics, in phase space, of the reduced state of the mirror and the field of one of the cavities. We determine occupation number and degree of squeezing of the equivalent squeezed thermal states used in order to effectively describe the reduced state of each subsystem. In Sec.~\ref{condition}, we track the changes undergone by these figures of merit when the optical part of the system or a proper ancilla is subject to conditioning measurements such as homodyne detection or projection onto the vacuum. Finally, Sec.~\ref{summary} summarizes our results.

\section{Introduction to the formalism} 
\label{formalism}

We consider an optical cavity with a movable, highly reflective end-mirror, coupled to the intracavity field by radiation pressure~\cite{law}. The Hamiltonian of the system reads (we take units such that $\hbar=1$ across the paper)
\begin{equation}
\label{coupling}
\hat{\cal H}=(\omega_c-G_0\hat{q}){\hat{a}}^\dag{\hat{a}}+\frac{\omega_m}{2}(\hat{p}^2+\hat{q}^2)-i{\cal E}(\hat{a}e^{i\omega_o{t}}-h.c.),
\end{equation}
where $\hat{a}$ and $\hat{a}^\dag$ are the cavity-field bosonic operators, $\omega_c$ its frequency and $\omega_m$ the frequency of the mirror, which is modelled as a harmonic oscillator with quadratures $\hat{q}=(\hat{m}^\dag+\hat{m})/\sqrt{2}$ and $\hat{p}=i(\hat{m}^\dag-\hat{m})/\sqrt{2}$ and bosonic operators $\hat{m},\,\hat{m}^\dag$. Moreover, ${\cal E}$ is the coupling rate between an external driving field (with frequency $\omega_o\simeq{\omega}_c$) and the cavity field and $G_0$ is the radiation-pressure coupling strength. Starting from $\hat{\cal H}$, Refs.~\cite{david,3Bpaper,ionjp} show how to obtain a set of linear Langevin equations useful to the reconstruction of the dynamics of the optomechanical system. This can be done by focusing our attention on the {\it fluctuations} of the relevant operators in the problem. The dynamics of the system is formally described by the integral matrix equation
\begin{equation}
\label{solution}
{\bf f}(t)=
e^{K_{}{t}}{\bf f}_{}(0)+\int^t_0{e}^{K_{}\tau}{\bf n}_{}(t-\tau)d\tau,
\end{equation} 
where ${\bf f}_{}(t)=(\delta{q}~\delta{p}~\delta{x}~\delta{y})^T$ is the vector of the quadrature fluctuations and ${\bf n}_{}(t)=(0~\xi(t)~\sqrt{2\kappa}\delta{x}^{in}~\sqrt{2\kappa}\delta{y}^{in})^T$ accounts for the noise entering the system. Here, $\xi(t)$ describes the Brownian motion of the mirror due to its coupling with a phononic bath at temperature $T$ and $\delta{x}^{in}=(\delta{a}^{in\dag}+\delta{a}^{in})/\sqrt{2}$ and $\delta{y}^{in}=i(\delta{a}^{in\dag}-\delta{a}^{in})/\sqrt{2}$ are the quadrature fluctuations of the input noise to the cavity, which has an amplitude decay rate $\kappa$. 
In the ordered basis given by $\{\delta{q},\delta{p},\delta{x},\delta{y}\}$, matrix $K$ reads
\begin{equation}
K=
\begin{pmatrix}
0&\omega_m&0&0\\
-\omega_m&-\gamma_m&G&0\\
0&0&-\kappa&\Delta\\
G&0&-\Delta&-\kappa
\end{pmatrix},
\end{equation}
where $G={G_0}{\cal E}/\sqrt{\kappa^2+\Delta^2}$ is a modified mirror-field coupling rate and $\Delta$ is the radiation-pressure affected system detuning~\cite{david,ionjp}. In the parameter regime where the eigenvalues of $K$ have negative real parts, Eq.~(\ref{solution}) can be solved at the steady state defined by $\lim_{t\rightarrow\infty}e^{Kt}=0$. The linear nature of the problem at hand guarantees the Gaussian-preserving character of the map defined by Eq.~(\ref{solution}). This allows us to focus on the entries of the {covariance matrix} ${\bm V}$ defined as $V_{ij}=\lim_{t\rightarrow\infty}\frac{1}{2}\langle\{{f}_{i}(t),f_{j}(t)\}\rangle$~\cite{eisertplenio}, where the expectation value is calculated over the mirror-field state. The explicit form of ${\bm V}$ is
\begin{equation}
\label{tipica}
{\bm V}=
\begin{pmatrix}
{\bm M}&{\bm C}\\
{\bm C}^T&{\bm F}
\end{pmatrix}=
\begin{pmatrix}
m_{11}&0&c_{11}&c_{12}\\
0&m_{22}&c_{21}&c_{22}\\
c_{11}&c_{21}&f_{11}&f_{12}\\
c_{12}&c_{22}&f_{12}&f_{22}
\end{pmatrix},
\end{equation}
where ${\bm M}\,({\bm F})$ is the $2\times{2}$ block matrix describing the local properties of the mirror (field) and ${\bm C}$ accounts for the mirror-field correlations. By assuming a large mechanical quality factor and a Ohmic spectral density of the background phononic bath~\cite{david,3Bpaper,benguria}, analytic expressions of the elements entering ${\bm V}$ can be easily found. They are however too lengthy to be reported here. As discussed above, Eq.~(\ref{solution}) does not affect the Gaussian nature of any input mirror-field state. 
In Refs.~\cite{david,3Bpaper}, this property has been used to show the entangling properties of the radiation pressure coupling in the presence of noise. Here, we concentrate on the phase-space properties of the reduced state of each subsystem.

\section{Reconstruction of the Wigner functions}
\label{Wigner}

In this Section we give an effective description of the local states of the mirror and the field in terms of equivalent linear optics transformations applied to proper input Gaussian states. An analogous study performed with respect to the joint two-mode state described by Eq.~(\ref{tipica}) is made difficult by the fact that $V$ describes a mixed state. Finding the corresponding linear-optics interferometer from which the state would be derived is, in general, a demanding task which goes beyond the scope of our investigation. A complete and self-contained account of the quantum correlations within $V$ has already been provided~\cite{david}. Here we concentrate on the reduced state of the mirror and the field as resulting from the radiation pressure-induced evolution. 
The inspection of Eq.~(\ref{tipica}) is quite revealing. Indeed, by looking at the form of ${\bm M}$, one notices that the mirror could be in a squeezed state with squeezing factor $\frac{1}{4}\ln(\frac{m_{22}}{m_{11}})$. In the phase space, such squeezing would necessarily be along the direction of one of the two mirror quadratures ({\it i.e.} the squeezing parameter must be real). On the other hand, the state of the field could be squeezed as well, but this has to occur along a direction that can be completely general. A formal account of such effects is given in Appendix A. Here, we make our expectations quantitative and concentrate on the behavior of the Wigner function associated with each reduced subsystem. These are related to the covariance matrices introduced in Eq.~(\ref{tipica}) by the expression 
\begin{equation}
\label{w}
W_J=\frac{e^{-\frac{1}{2}f_{J}{\bm J}^{-1}f^T_{J}}}{\pi\sqrt{\det{\bm J}}},
\end{equation}
where $J=M$ or $F$. Here, $f_{M}=(\delta{\tilde q}~\delta{\tilde p})$, $f_{F}=(\delta{\tilde x}~\delta{\tilde y})$ are complex vectors for the mirror and field quadrature variables. For the mirror, we have taken 
\begin{figure}[t]
\psfig{figure=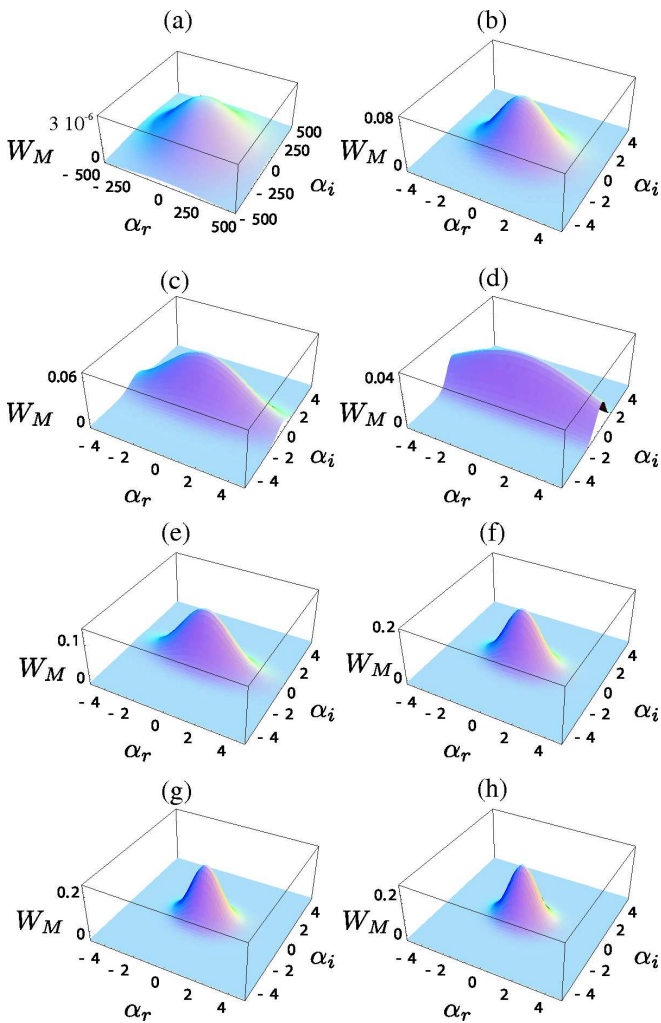,width=\columnwidth}
\caption{(Color online) Wigner function associated with the mirror state plotted against the rescaled detuning $\Delta=j\omega_m/10~~(j\in{\mathbb Z})$. Each panel corresponds to a set value of $j$. We have taken $j=0~{\bf (a)}$, $4~{\bf (b)}$, $7~{\bf (c)}$, $10~{\bf (d)}$, $13~{\bf (e)}$, $16~{\bf (f)}$, $19~{\bf (g)}$ and $20~{\bf (h)}$. Throughout the paper we have used the following set of parameters: $\omega_m/2\pi=10$MHz, $\omega_o\simeq\omega_c/2\pi=3.7\times10^{14}$Hz, ${\cal E}=6\times10^{12}$Hz, $G_0\simeq1400$Hz, $T=0.4$K, and a cavity of finesse $10^4$ that is $1$mm long.}
\label{wignermirror}
\end{figure}
$\delta\tilde{q}=(\alpha+\alpha^*)/\sqrt{2}$ and $\delta\tilde{p}=i(\alpha^*-\alpha)/\sqrt{2}$ with $\alpha=\alpha_r+i\alpha_i$ and the field quadrature variables $\delta{\tilde x}$ and $\delta{\tilde y}$ have been analogously defined. With these choices, $W_M(\alpha_r,\alpha_i)$ turns out to be a function of the effective detuning $\Delta$. In Fig.~\ref{wignermirror} we show such a dependence by setting $\Delta$ and studying $W_M(\alpha_r,\alpha_i)$ against $\alpha_r$ and $\alpha_i$. For a resonant interaction [panel {\bf (a)}], $W_M(\alpha_r,\alpha_i)$ is an isotropic Gaussian with a large width (notice the range of values of $\alpha_{r,i}$), which is indicative of a thermal state with large mean occupation number. 
As $\Delta$ grows, the variance of $W_M(\alpha_r,\alpha_i)$ along the $\alpha_r$ direction becomes sensibly larger than that along $\alpha_i$, thus showing a $\Delta-$dependent squeezing effect. However, the direction of squeezing remains fixed, thus indicating that the squeezing parameter is always real and negative.
The width of the Wigner function, which gives information on the effective temperature of the mirror, is considerably reduced in a large range of values of $\Delta$, thus revealing a strong cooling effect, even for moderate detuning between the cavity and the external field. 
This effect has been extensively studied in a series of recent experiments~\cite{cool}. For large values of $\Delta$, the state of the mirror goes back to a {thermal} behavior, with a reduced mean occupation number, which slowly increases toward its initial value as $\Delta$ is further increased. In fact, with a growing detuning, the effective coupling rate between mirror and field decreases as less input power couples into the far off-resonant cavity, thus driving the mirror towards its unperturbed initial thermal state. 

In order to make these considerations more quantitative, we proceed as follows. The most general single-mode Gaussian state of a boson of frequency $\omega$ is given by the squeezed thermal state
\begin{equation}
\rho_{gen}=Z(\beta)\hat{S}(re^{i\varphi}){\rm e}^{-\frac{\beta}{2}(2\hat{n}+1)}\hat{S}^{\dag}(re^{i\varphi}),
\end{equation}
 where $Z(\beta)=e^{\beta/2}-e^{-\beta/2}$ is the partition function of a thermal state having effective temperature $\beta^{-1}=k_bT/\omega$ ($k_b$ is the Boltzmann constant), $\hat{S}(re^{i\varphi})$ is the single-mode squeezing operator with complex amplitude $re^{i\varphi}$ defined in Appendix A and $\hat{n}$ is the bosonic number operator. Here, we have neglected the possibility of a displacement of the state in the phase-space as the quadrature operators we are considering refer to the zero-mean fluctuations of both the field and mirror.  
In order to quantify the similarity between the mirror state and $\rho_{gen}$, we could compare the behavior of the associated Wigner functions.
However, as we are dealing with Gaussian states, a much more convenient comparison is performed in terms of covariance matrices~\cite{commentocomparison}. We use $s_{J}e^{i\varphi_{J}}$ and $\nbar_{J}+1/2$ to indicate the effective squeezing factor and thermal variance of mode $J=M,F$. In this way, a full characterization of each reduced state is obtained by numerically solving the matrix equation
\begin{equation}
\label{risolvospecchio}
{\bm J}={\bm v}^{J}_{\cal S}\equiv{\cal S}(s_J,\varphi_J)
(\nbar_J+\frac{1}{2})\openone_2
{\cal S}(s_J,\varphi_J),
\end{equation}
for the unknown effective parameters $\nbar_J\,s_J$ and $\varphi_J$, where ${\cal S}(s_J,\varphi_J)={\cal S}^T(s,\varphi)$ is the symplectic transformation corresponding to single-mode squeezing of amplitude $s_Je^{i\varphi_J}$ [cfr. Eq.~(\ref{squeeze})] and $(\nbar_J+\frac{1}{2})\openone_2$ is the covariance matrix of a single-mode thermal state ($\openone_2$ is the $2\times{2}$ unit matrix). Any explicit dependence of $\nbar_{J}$ and $s_J$ on $\Delta$ has been omitted for convenience of notation. 
\begin{figure}[b]
\psfig{figure=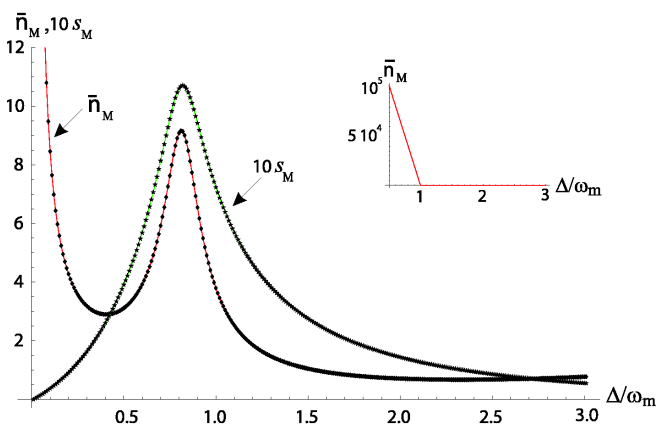,width=\columnwidth}
\caption{(Color online) Effective squeezing factor $s_M$ (rescaled so as to improve its visibility) and average phonon number $\nbar_{M}$ against the dimensionless detuning $\Delta/\omega_M$.
The inset shows $\nbar_M$ with an enlarged vertical axis so as to provide evidence of the considerable cooling experienced by the mirror. We find $\varphi_M=\pi$, regardless of $\Delta$.}
\label{confronti}
\end{figure}
The solution of Eq.~(\ref{risolvospecchio}) for $J=M$ leads to the behaviors shown in Fig.~\ref{confronti}. As expected, we find $\varphi_M=\pi$, in line with our conjecture on the antisqueezing of the mirror state. On the other hand, $\nbar_M$ and $s_M$ are not monotonous against $\Delta$. In particular, the effective mean occupation number is peaked in correspondence with the maximum of $s_M$. Although our numerical approach allows for the immediate visualization of the results, it is possible to give an analytic expression for both the degree of squeezing and the mean occupation number. We refer to Appendix B for full analytical details.
\begin{figure}[b]
{\bf (a)}
{\psfig{figure=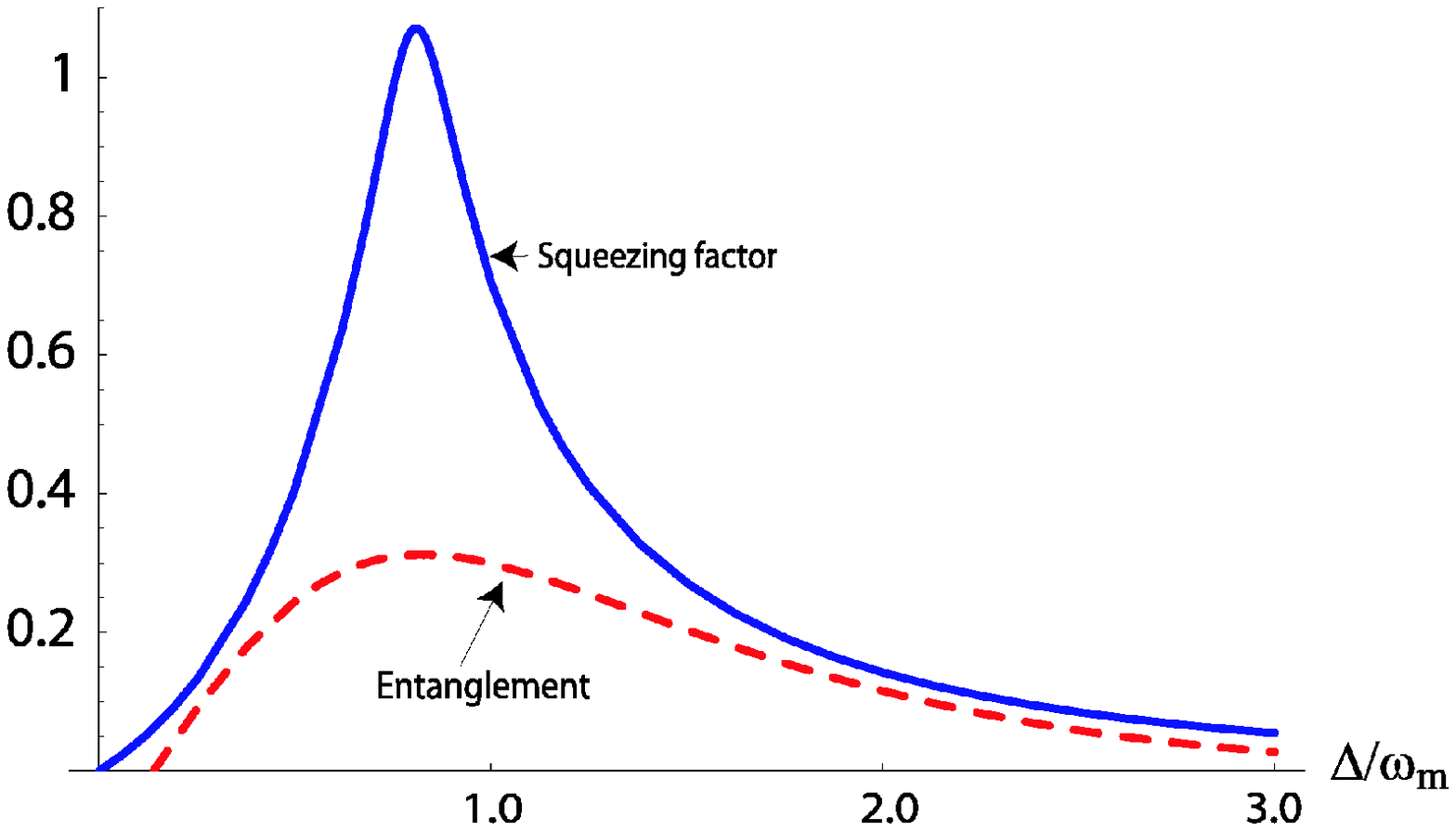,width=\columnwidth}\\
{\bf (b)}
\psfig{figure=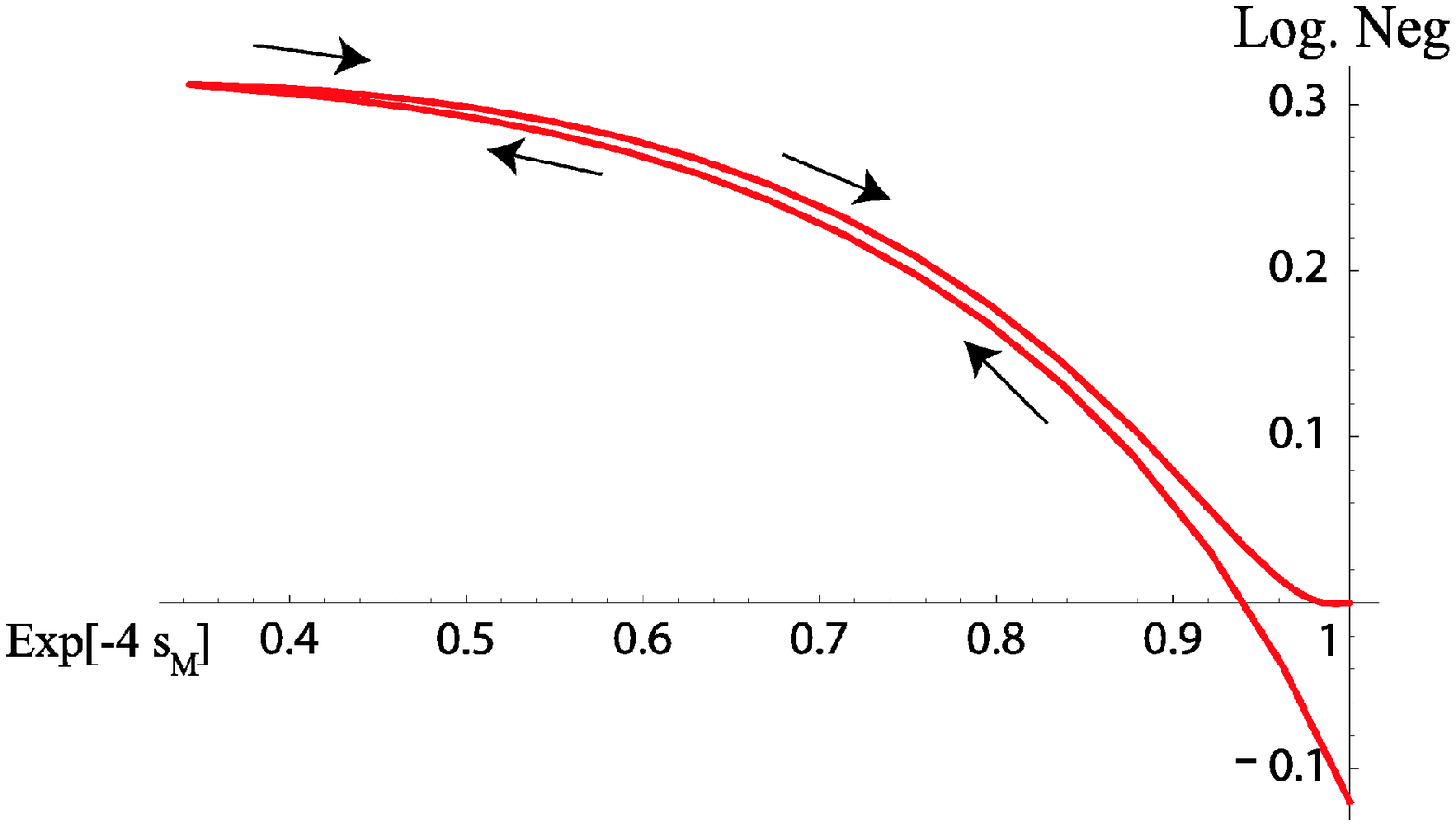,width=\columnwidth}}
\caption{(Color online) {\bf (a)}: Modulus of the effective squeezing factor $s_M$ and entanglement in the mirror-cavity system plotted against $\Delta/\omega_m$. The two curves are peaked at the same detuning. {\bf (b)}: Logarithmic negativity in the mirror-field system~\cite{david,3Bpaper} shown against $e^{-4 s_M}$. The horizontal axis is truncated at $s_M\simeq{0.3}$ for clarity. 
}
\label{aggiunto}
\end{figure}

A natural question arises, at this point: How is the effective degree of squeezing related to the optomechanical entanglement that is known to occur in the conditions at hand? In order to provide a quantitative answer, we have calculated the logarithmic negativity of the mirror-field system~\cite{david,3Bpaper} as a function of $e^{-4s_M}$, for $s_M$ going from $0$ to its maximum $\simeq{1.1}$ and back to zero again. Two different values of entanglement will be associated with the same degree of squeezing $s_M$, depending on the way the detuning has been changed. This is an effect of the asymmetry, with respect to its maximum, of the curve showing entanglement against detuning [Fig.~\ref{aggiunto} {\bf (a)}]. Such considerations are confirmed by the incomplete hysteresis loop-like curve displayed in Fig.~\ref{aggiunto} {\bf (b)}. In each branch of the hysteresis loop, entanglement is a monotonic (increasing or decreasing) function of the squeezing. The qualitative interpretation of our results goes along the following lines. In the linearized regime where Eq.~(\ref{solution}) is valid, the radiation-pressure coupling term $-G_0\hat{a}^\dag\hat{a}\hat{q}$ in $\hat{\cal H}$ includes both energy-preserving and non-preserving terms~\cite{io}. On their own, the energy preserving terms, having the form $\hat{a}\hat{m}^\dag+h.c.$, would generate a beam-splitting operation involving mirror and cavity field. The non-preserving term $\hat{a}^\dag\hat{m}^\dag+h.c.$ is the generator of a two-mode squeezing transformation~\cite{barnettradmore}. Under this viewpoint, the full optomechanical dynamics encompassed by Eq.~(\ref{coupling}) and without the inclusion of noise effects, can be interpreted in terms of an equivalent interferometric setting where two modes impinge at a beam splitter. It is well-known that, in order for the output of such a device to show entanglement, a certain degree of non-classicality (such as squeezing) should be present at the input. Therefore, maximum entanglement is achieved at the conditions that, dynamically, optimize the squeezing of the mirror mode. Clearly, the true physical situation at hand is somehow different from this intuitive picture as we deal with a stationary state that is affected by losses and noise. The above discussion, however, is sufficient to get an intuition of the physical process at hand. 

\begin{figure}[t]
\psfig{figure=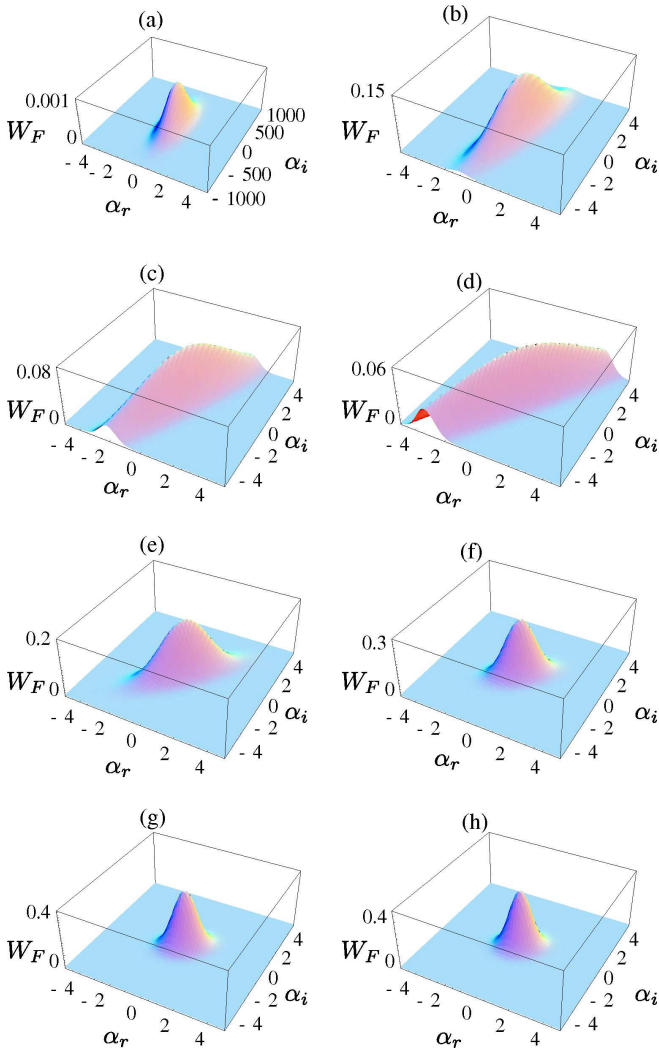,width=\columnwidth}
\caption{(Color online) Wigner function associated with the cavity field plotted against $\Delta=j\omega_m$. He have taken $j=0,~0.4,~0.7,~1,~1.3,~1.6,~1.9,~2$ in going from {\bf (a)} to {\bf (h)}, respectively.}
\label{wignerfield}
\end{figure}

\begin{figure}[ht]
\psfig{figure=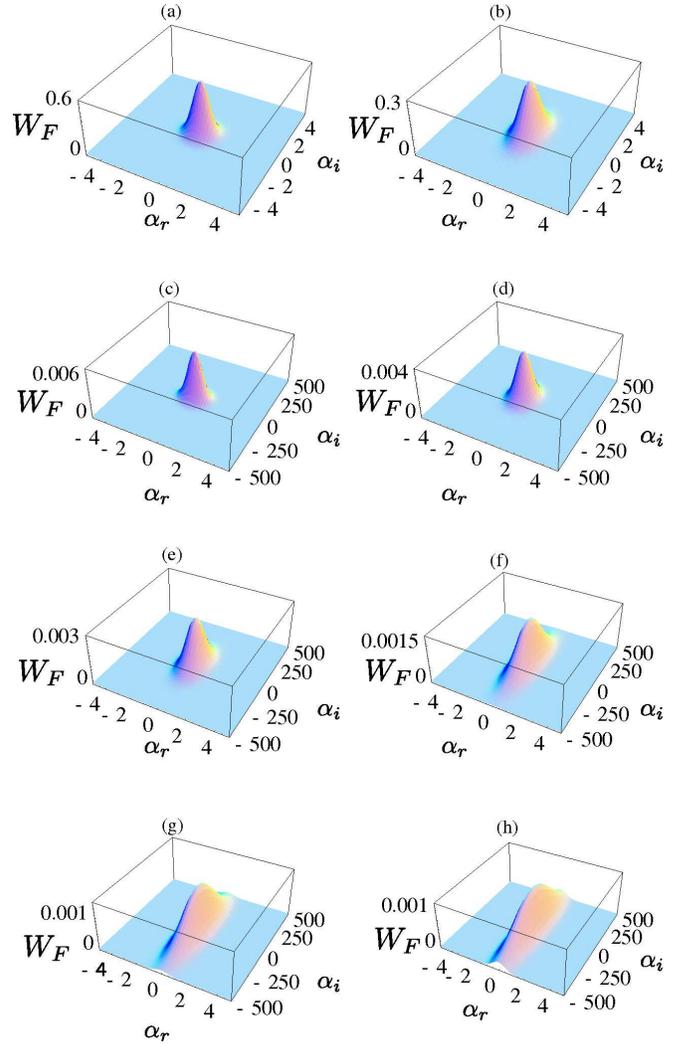,width=\columnwidth}
\caption{(Color online) Wigner function associated with the cavity field studied against the parameter $\chi$ that tunes the optomechanical coupling. We have taken $\chi=0.01,~0.04,~0.4,~0.5,~0.6,~0.8,~0.9,1$ in going from {\bf (a)} to {\bf (h)}, respectively.}
\label{wignerfieldcontrocoupling}
\end{figure}

Let us now shift our attention to the cavity field state and consider the Wigner function $W_{F}(\alpha_r,\alpha_i)$ associated with its reduced state. This is easily calculated by means of Eq.~(\ref{w}) with $J=F$ and its behavior is shown in Figs.~\ref{wignerfield} for set values of $\Delta$. It is interesting to note that squeezing appears in the field state even at zero detuning, as witnessed by the evident deformation of the bidimensional Gaussian in Fig.~\ref{wignerfield} {\bf (a)}. This is in contrast with the behavior of the mirror state, which requires $\Delta\neq{0}$ in order to be squeezed. Therefore, as it will be clarified later, one cannot relate the squeezing of the field to the entanglement in the joint cavity-mirror state.  The cavity field squeezing appears to be a simple consequence of the coupling to the mirror. Indeed, we have studied the evolution of the field squeezing at $\Delta=0$ and by replacing the coupling parameter $G$ with $g=\chi{G}$, where $\chi\in[0,1]$. This accounts for a resonant cavity-pump configuration and a tunable power of the cavity-driving field. The shape of $W_{F}(\alpha_r,\alpha_i)$ for proper values of $\chi$ is shown in Figs.~\ref{wignerfieldcontrocoupling}, where we see the rapid 
appearance of squeezing (in panel {\bf (c)} we had to enlarge the range of $\alpha_i$ up to $[-500,500]$ in order for the Wigner function to look isotropic), more pronounced as the scaling factor increases. The sudden squeezing of the cavity field state, which is absent in the mirror state, may be due to the different initial conditions assumed for the two subsystems. The explicitly mixed nature of the mirror state (due to the thermal background of phononic modes) requires the enhancement of the nonlinear character of the interaction with the field in order to exhibit significant squeezing. This is effectively achieved via a non-zero detuning. On the other hand, as the cavity field is prepared in a coherent state, even a resonant interaction significantly affects its state.  

\begin{figure}[b]
{\bf (a)}\hskip3cm{\bf (b)}
\psfig{figure=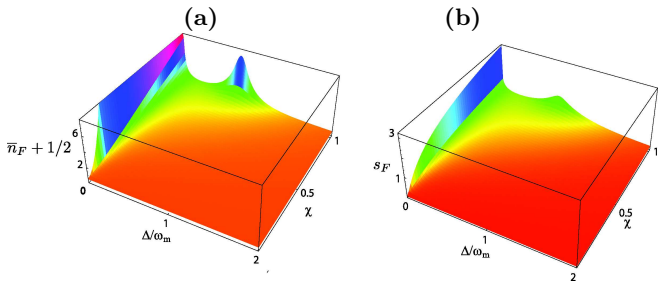,width=\columnwidth}
\caption{(Color online) Effective variance {\bf (a)} and degree of squeezing {\bf (b)} of the equivalent field state plotted against $\Delta/\omega_m$ and $\chi$.}
\label{solido}
\end{figure}

By following the same line as in the case of the mirror, 
we can calculate the effective mean occupation number and degree of squeezing of the field state against the optomechanical coupling and the detuning, so as to gather a complete picture (see Fig.~\ref{solido}). It is easy to show that 
\begin{equation}
\begin{aligned}
\nbar_F=-\frac{1}{2}+\sqrt{\det{\bm F}},~~~~~s_F=\frac{1}{2}{\rm arccosh}\left(\frac{{\rm Tr}{\bm F}}{2\sqrt{\det{\bm F}}}\right),
\end{aligned}
\end{equation}
which can be easily calculated by using Eq.~(\ref{tipica}) and the approaches of Refs.~\cite{david,3Bpaper}. Any resulting expression is however too lengthy to be reported here and we summarize our findings in Fig.~\ref{solido}. First, $\nbar_F$ at $\Delta=0$ appears to be a quadratically-increasing function of $\chi$, as we have checked by finding the best fit function corresponding to this configuration [cfr. panel {\bf (a)}]. The effective squeezing, on the other hand, starts from zero as a concave function, soon becomes convex and grows as $\sqrt{\chi}$, as shown in Fig.~\ref{solido} {\bf (b)}. This proves that even at $\Delta=0$ squeezing of the field state should be expected. On the other hand, as previously commented, we have checked that no squeezing appears in the mirror state at resonance.
At $\Delta=0$, tuning $\chi$ simply determines the isotropic increase of the width of $W_M(\alpha_r,\alpha_i)$. 

We now address a second interesting point: different from the case of the mirror state, $W_F(\alpha_r,\alpha_i)$ at $\chi=1$ rotates (in the phase space) with the degree of squeezing, which implies that the squeezing factor of the field state is complex and provides additional information on the mechanism for setting entanglement within the system.
In fact, by using again the equivalent interferometric description depicted before, one can conclude that along with the amount of squeezing of the mirror state, an important role is played by the relative direction of squeezing of the two subsystems states. While the state of the mirror is squeezed along a fixed direction in phase space, the direction of squeezing of the field changes with the detuning as shown in Fig.~\ref{wignerfield}, thus affecting the amount of entanglement in the output modes. In this picture, we expect the detuning to enter in a critical way in the splitting ratio of the effective beam splitter that superimposes the two modes. It is important to note that a study including larger values of $\chi$ is not possible unless the set of parameters we are using is adjusted. Indeed, larger values of $\chi$ soon put the system in a regime of instability where a stationary solution to the Langevin equations is no longer possible. As we would like to keep the level of technicalities related to the search for stability regions away from our discussion, we refer to Refs.~\cite{david,3Bpaper} for a detailed description of this issue. 

\section{Conditional measurements}
\label{condition}

We now address the second relevant point of our study, namely the search for measurements on the accessible part of the system that allow to condition the state of the mirror. For instance, by taking the degree of squeezing of the reduced state of the mirror as an indicator of its non-classicality, is there a way to push it closer to the vacuum limit? Here we consider measurements performed over the field interacting with the mechanical mirror as well as an ancillary mode that is superimposed to the field state at a beam splitter. One can in fact find the covariance matrix of the system comprising the mirror and the extra-cavity field. It is related to the intra-cavity field studied so far via the well-known Collett-Gardiner input-output relations~\cite{inpout}. A detailed calculation (see Ref.~\cite{3Bpaper}) reveals that, with a proper temporal/frequency-filtering of the quadrature values of the optomechanical system, ${\bf V}$ in Eq.~(\ref{tipica}) can be reconstructed from the (experimentally determined) output covariance matrix $V_{out}$ as $V\propto(V_{out}-\openone/2)$. In Ref.~\cite{vitaliout} it was shown that this technique can also be used to have robust entanglement between the output mode and the mechanical mirror. More recently, the pulsed optomechanical paradigm has been proposed as an effective mean to reconstruct the state of a mechanical device~\cite{PulsedTheo}. A first proof-of-principle demonstration has been reported recently~\cite{PulsedExp}.

In any case, the linearity of the relations at hand allows to work directly with $V$. We restrict our attention to Gaussian measurements, which preserve the Gaussian character of the state being measured. We explicitly consider projective measurements of a field mode onto the vacuum and homodyne measurements.
The changes at the level of second moments of the quadrature operators after either a projection onto the vacuum or a homodyne measurement can be given in terms of Schur complements~\cite{eisert}. Let us assume we have the following covariance matrix of $n_1+n_2$ modes
\begin{equation}
\label{teoria}
{\bf V}_{ex}=
\begin{pmatrix}
{\bf A}&{\bf C}\\
{\bf C}^T&{\bf B}
\end{pmatrix},
\end{equation}
where ${\bf B}$ gives the local variances of quadrature operators in a system of $n_2$ modes to be measured, ${\bf A}$ describes the analogous quantities for the remaining $n_1$ modes and ${\bf C}$ accounts for the correlations between the $n_1$- and $n_2$-mode systems. For a {homodyne measurement} over the $n_2$ modes, the ``updated'' covariance matrix of the remaining $n_1$ modes ({\it i.e.} the covariance matrix of the system after the measurement has been performed) is given by the Schur complement
\begin{equation}
\label{updatedhomo}
{\bf A}'_{hom}={\bf A}-{\bf C}[{\bm \pi}{\bf B}{\bm \pi}]^{-1}_{mpi}{\bf C}^T,
\end{equation}
where ${\bm \pi}\!=\!\oplus^{n_2}_{j=1}\begin{pmatrix}1&0\\0&0\end{pmatrix}$ and $[\cdot]^{-1}_{mpi}$ stands for the Moore-Penrose pseudo-inverse of a matrix~\cite{eisert,eisertplenio}. On the other hand, if the $n_2$ modes are projected onto the vacuum, the updated covariance matrix of the remaining subsystem is 
\begin{equation}
\label{updatedvacu}
{\bf A}'_{vac}={\bf A}-{\bf C}({\bf B}+\one_{})^{-1}{\bf C}^T,
\end{equation}
where $\one$ is the $2n_2\times{2n_2}$ identity matrix and the standard inverse of a matrix is used. While ${\bf C}$ is a $2n_1\times{2n_2}$ matrix, both ${\bm \pi}{\bf B}{\bm \pi}$ and  ${\bf B}+\one$ have dimension $2n_2\times{2n_2}$. The term to be subtracted to the {pre-measurement} covariance matrix ${\bf A}$ is always a $2n_1\times{2n_1}$ matrix, as it should be.
We first consider a measurement to be operated directly on the cavity field.
For the case at hand, $n_1=n_2=1$, ${\bf A}={\bf M}$, ${\bf B}={\bf F}$ and the calculation of the updated covariance matrix of the mirror is quite straightforward~\cite{footnote}. Through  an equation analogous to Eq.~(\ref{risolvospecchio}) but involving, this time, ${\bf M}'_{hom,vac}$, we easily get information about the effective mean occupation number and squeezing of the conditional state of the mirror. 

\begin{figure}[b]
{\bf (a)}
\psfig{figure=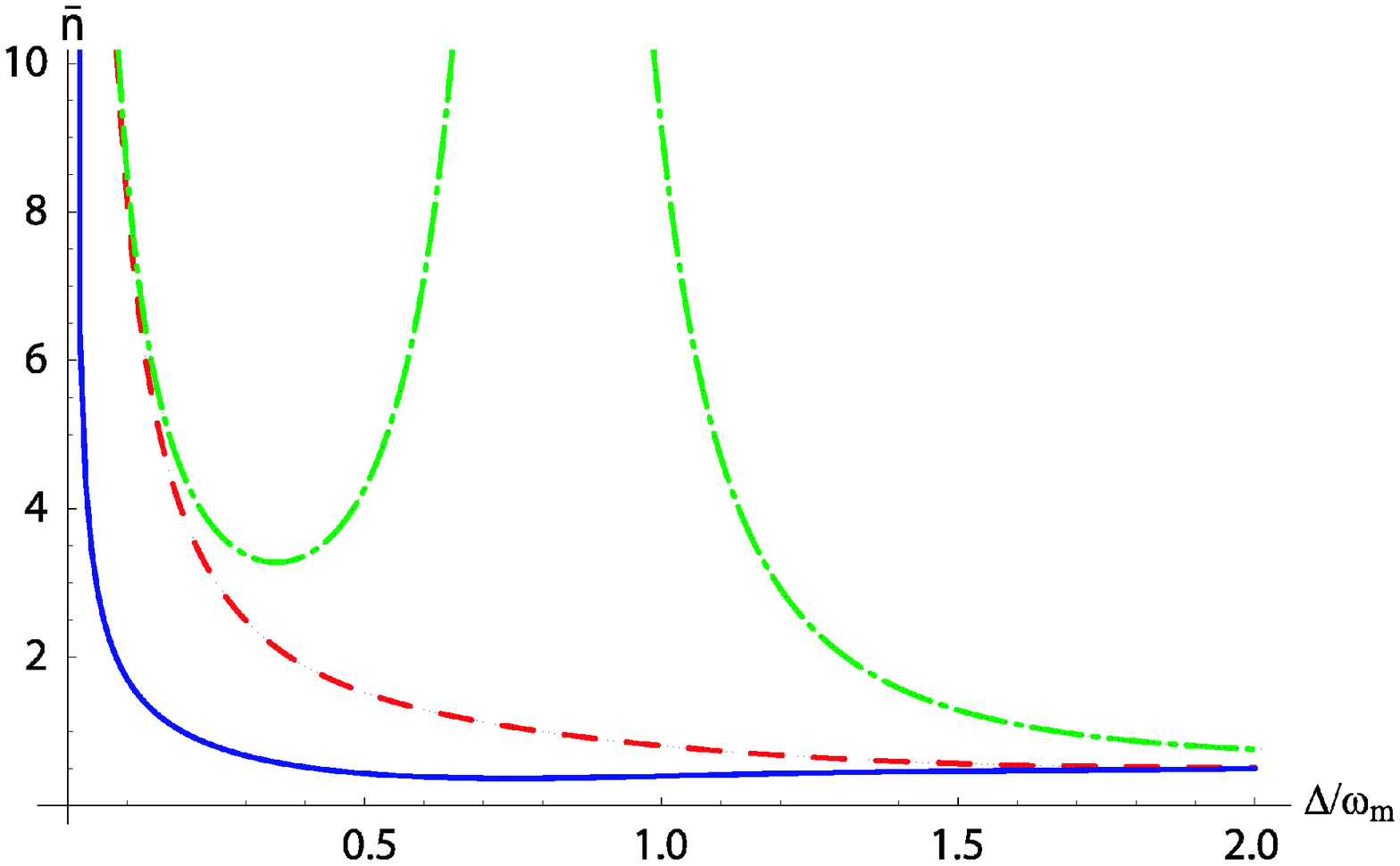,width=\columnwidth}\\
{\bf (b)}
\psfig{figure=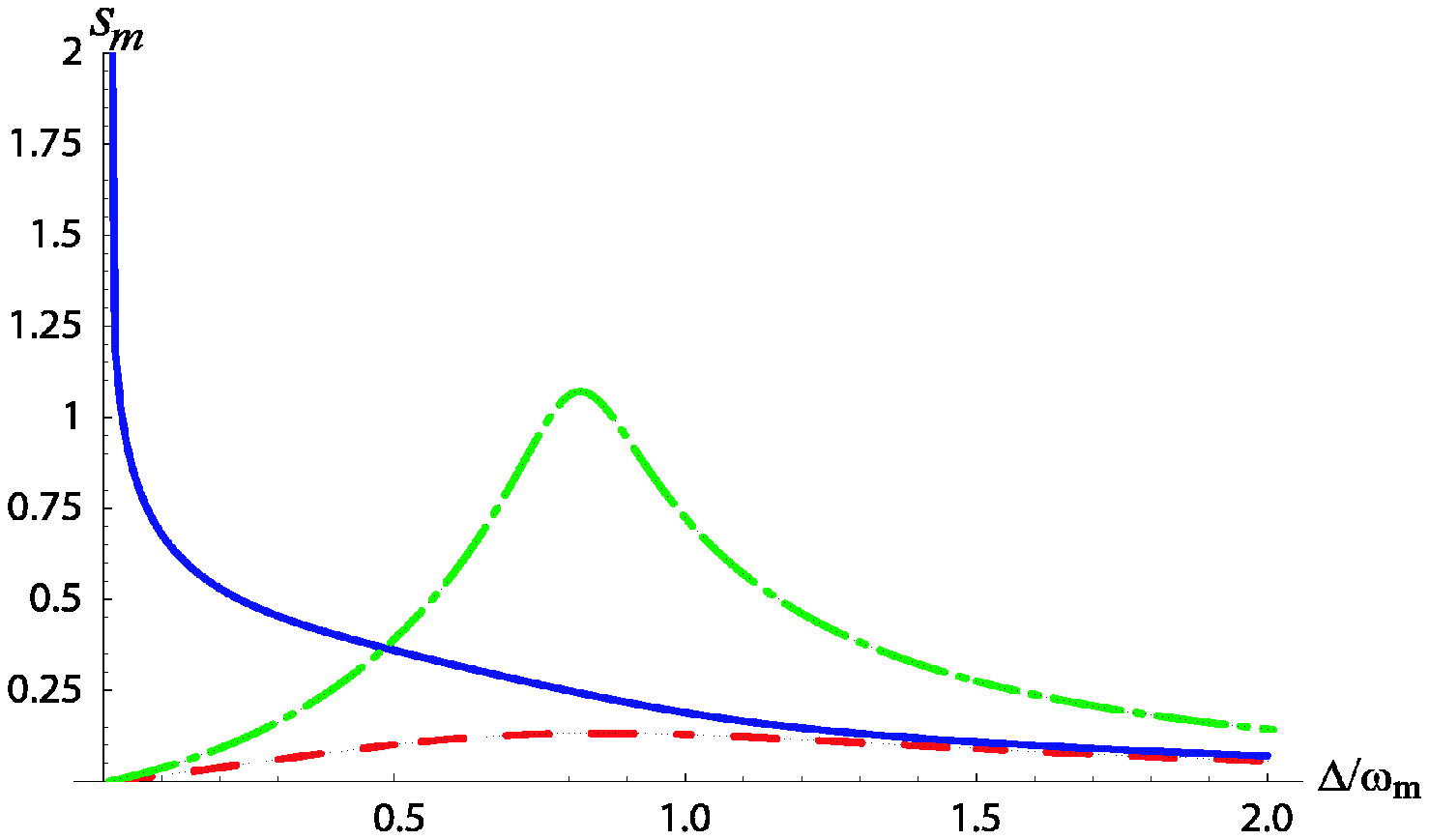,width=\columnwidth}
\caption{(Color online) Panel {\bf (a)} shows the effective mean occupation number of the mirror  state against $\Delta/\omega_m$. Panel {\bf (b)} reports the effective squeezing. Dot-dashed lines show the properties of the unconditioned mirror state, dashed lines are for the homodyne-measurement  case and solid lines are for the vacuum-projection case.}
\label{confrontineff}
\end{figure}

\begin{figure}[t]
\psfig{figure=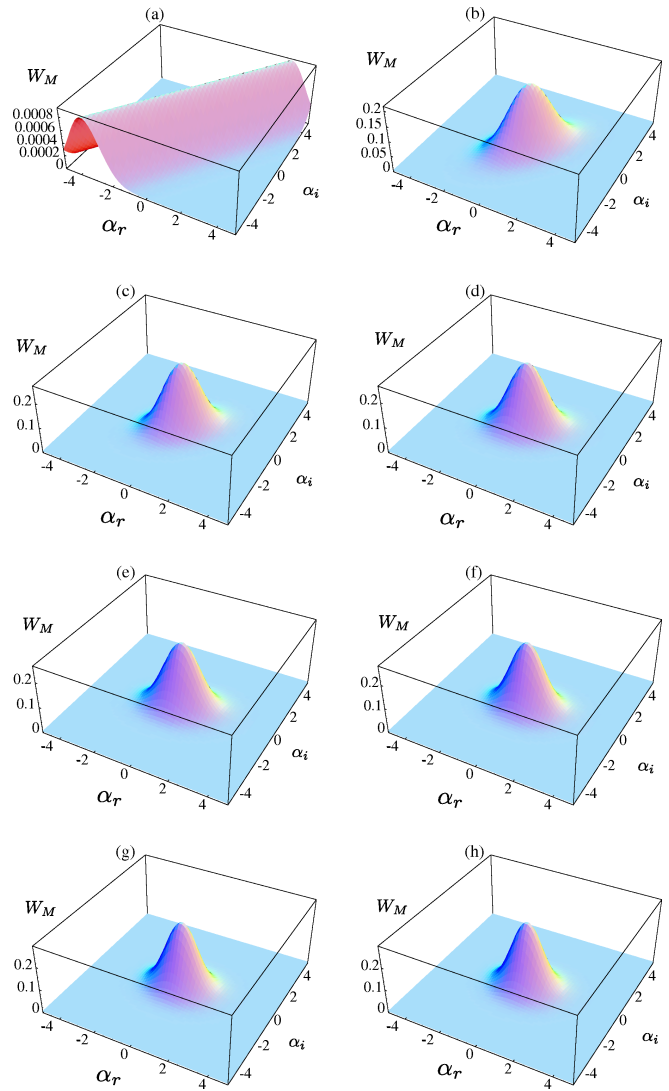,width=\columnwidth}
\caption{(Color online) Snapshots of the evolution of the Wigner function associated with the mirror state against the detuning $\Delta$ [which increases from 0 to $10\omega_m$ in going from {\bf (a)} to {\bf (h)}] for the case of projection onto vacuum of the cavity field.}
\label{wignersequenceproj}
\end{figure}

The comparison between the case of vacuum-projected (solid line), homodyne-measured (dashed line) and non-conditioned (dot-dashed line) mirror state is shown in Fig.~\ref{confrontineff} {\bf (a)}.
Effective cooling of the mirror is achieved, for both the homodyne measurement and the projection onto vacuum. At $\Delta\simeq{0.8}\omega_m$, a reduction of $\nbar_M$ of almost one order of magnitude with respect to the un-conditioned case is achieved when a vacuum-projection is implemented.  
This is accompanied by the raising of the effective squeezing parameter at small detuning, as shown in Fig.~\ref{confrontineff} {\bf (b)}.
In this sense the projection onto vacuum, being far from a ``classical'' measurement such as homodyning, pushes the mirror state towards non-classicality. 
Cooling and squeezing effects are both accompanied by a rotation of the Wigner function associated with the state of the mirror (see Fig.~\ref{wignersequenceproj}, where the large initial squeezing and its decrease to zero are clearly evident).

Although effective in pushing the mirror towards a more non-classical state, the strategy used above obviously cannot be used to affect the optomechanical entanglement. In order to achieve this task, we have to consider an ancillary mode that effectively introduces an additional control. 
We thus consider a second field, superimposed to the one studied so far at a beam splitter with a tunable splitting ratio $\theta$. The ancilla is initially prepared in its vacuum state and, consistently with the analysis above, projections onto vacuum are considered. The mixture of the cavity field with the ancilla can be described formally by considering the analogy between injecting a field into a cavity and a beam splitting operation $\hat{B}_s=\exp[\theta(\hat{a}^\dag\hat{b}-h.c.)]$, where $\hat{b}$ is the annihilation operator of the ancilla and $\theta$ is the splitting ratio.


\begin{figure}[ht]
\psfig{figure=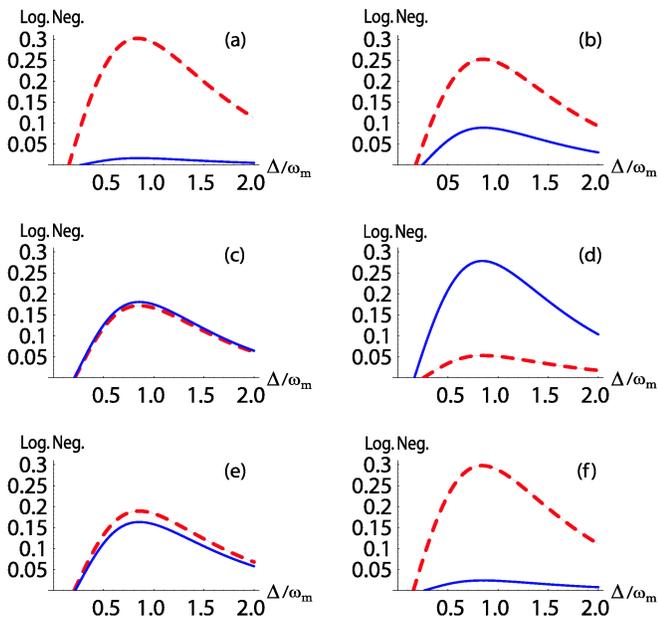,width=\columnwidth}
\caption{(Color online) Entanglement in the mirror-cavity (dashed line) and mirror-ancilla system (solid line) against the detuning parameter and as $\theta$ is changed from $0$ to $\simeq{\pi}$. Entanglement is created in one subsystem at the expenses of the other one. In panel {\bf c} ({\bf (e)}), $\theta\simeq{\pi/4}$ ($\theta\simeq{3}\pi/4$), while panel {\bf (d)} shows the case of $\theta$ close to $\pi/2$.
}
\label{tripartito}
\end{figure}

In order to correctly understand what has to be expected from this thought-experiment, we have to analyze the dynamics of the entanglement within the tripartite system comprising the mirror and the two fields. This is easily done by considering the covariance matrix of the system after the beam splitting operation superimposing the cavity field and the ancilla and evaluating the logarithmic negativity of each of bipartition extracted from the tripartite system. The analysis can be performed analytically to some extent and, in principle, complete expressions for the logarithmic negativity of every reduced bipartite state can be given, although they are very complicated. We find that no entanglement can be established between the cavity field and the ancilla, regardless of the splitting ratio used for the beam splitter and the detuning in the cavity-mirror system. The two field modes remain separable. On the other hand, there is an interplay between the entanglement in the mirror-cavity field and in the mirror-ancilla subsystem. In general, both can be entangled, thus revealing a {\it one-mode biseparable} (or two-way entangled) nature of the trimodal state we are analysing, where the classification given in~\cite{giedke} is used. This is somehow complementary to the three-body scheme that has been considered in Ref.~\cite{3Bpaper}, where a genuine tripartite entangled state is created. The quantitative results relative to the two-way entangled state we get here are shown in Fig.~\ref{tripartito}, where we present the logarithmic negativity~\cite{ent} between mirror and cavity field (dashed line) and between mirror and ancillary field (solid line) against the detuning, for different values of the splitting ratio $\theta$. The entanglement initially present in the cavity-mirror system is {\it poured} into the ancilla-mirror one as $\theta$ goes from $0$ to $\pi/2$, when the mirror and the cavity field appear to be separable, in favour of the quantum correlations between ancillary field and mirror. By increasing $\theta$, the specular situation is obtained with the entanglement in the mirror-cavity system being eventually restored. At $\theta=\pi/4$ (and $\theta=3\pi/4$) a symmetric one-mode biseparable state is found: the entanglement in the two subsystems is the same.

However, the most interesting effects are at the level of the mirror state. In terms of equivalent squeezed thermal states, the interaction with the ancillary mode and its subsequent projection onto the vacuum cools down the mirror state. This can be put in correspondence with the entanglement distributed in the trimodal state. At $\theta=0$, the mirror and the ancilla are not entangled at the beam splitter. The situation goes exactly as in the case without ancilla and we expect a behavior which replicates the one corresponding to a {\it no-measurement case}. However, as soon as $\theta\neq{0}$, measuring the state of the ancilla means, effectively, reducing the temperature of the mirror, which reaches a minimum at $\Delta\simeq{\omega_m}$ and $\theta=\pi/2$, when all the entanglement in the trimodal state is in the mirror-ancilla subsystem.  
It is possible to provide an analogous study of the behaviour of the effective squeezing in the conditioned state. We mention that the squeezing function of the conditioned state will initially ($\theta=0$) coincide with the squeezing function of the unconditioned state and then evolves toward a situation of large squeezing at small detuning. The behavior becomes exactly the same as in Fig.~\ref{confrontineff} with $\theta=\pi/2$.

\section{Conclusions}
\label{summary}

We have studied the phase-space behavior of an optomechanical system comprising a movable mirror coupled to the field of a driven optical cavity, a device which is currently of enormous experimental interest. The advantages inherent in our phase-space approach are quite evident: By modelling the reduced state of each subsystem (in the presence of the relevant noise mechanisms) as a squeezed thermal state, we have shed light onto the optomechanical dynamics originating from their radiation-pressure coupling in a clear way, which also offered quite interesting insight. In fact, we have revealed that the optical and mechanical subsystems behave in qualitatively different ways: The cavity field is squeezed regardless of the cavity-pump detuning, but is also rotated in the phase space, a feature that misses entirely in the mechanical dynamics. We have studied the behavior of the squeezing of the mechanical system, its thermal occupation number and entanglement with the optical field against the most relevant parameters of the model. 

The significance of the analysis thus performed is also revealed by the fact that the information gained through this approach has allowed the design of a strategy for the conditional cooling of the mirror state via measurements performed on the cavity field. In this respect, we have shown that, by postselecting the vacuum state of the cavity field, one can generate an effective measurement-induced non-linearity sufficient to considerably modify the dynamics of the mirror, inducing additional cooling and squeezing-enhancing effects. 

 Our study is a step forward towards the complete understanding of the evolution of an open optomechanical system, whose dynamical features can be significantly different from the unitary dynamics of its noiseless counterpart~\cite{io}. In the quest for the coherent use of massive micro-mechanical systems operating at the quantum limit, it is mandatory to achieve the full control of their coupling to light, which serves as the ideal ancilla for quantum state engineering, information processing and state revelation. By suggesting conditional processes that are able to effectively manipulate the state of an inaccessible mechanical system, our work contributes to such an important goal.

\acknowledgments
We thank M. Aspelmeyer and H. Jeong for invaluable discussions at the various stages of preparation of this work. M.P. is supported by the UK EPSRC (EP/G004579/1). S.Gr. is the recipient of a Marie Curie Fellowship from the European Commission. S.G. is funded by the European Research Council (ERC COMEDIA). 

\renewcommand{\theequation}{A-\arabic{equation}}
\setcounter{equation}{0}  
\section*{APPENDIX A}  

In this Appendix, we give a technical description of the general effect that squeezing has on the covariance matrix of a single bosonic system prepared in a Gaussian state. Let us consider the single-mode squeezing operation $\hat{S}(se^{i\varphi})=\exp[{\frac{s}{2}e^{i\varphi}\hat{b}^{\dag2}-h.c.}]$ of squeezing parameter $se^{i\varphi}$ (here $\hat{b}^\dag$ is the creation operator of a general bosonic mode)~\cite{barnettradmore}. This changes a covariance matrix 
${\bm v}=\text{Diag}[v_0,v_1]$ (which is general enough to represent, under proper conditions, a coherent or a thermal state) via the corresponding symplectic transformation
\begin{equation}
\label{squeeze}
{\cal S}(s,\varphi)=
\begin{pmatrix}
\cosh{s}-\cos\varphi\sinh{s}&-\sin\varphi\sinh{s}\\
-\sin\varphi\sinh{s}&\cosh{s}+\cos\varphi\sinh{s}
\end{pmatrix}
\end{equation}
as ${\cal S}^T(s,\varphi){\bm v}{\cal S}(s,\varphi)$.
For $\varphi=0$ or $\pi$, ${\bm v}_{\cal S}$ (corresponding to squeezing and antisqueezing of the bosonic mode, respectively) reduces to a diagonal form. Therefore, a real squeezing factor (either positive or negative) corresponds to a diagonal covariance matrix of the squeezed bosonic mode. This is the case for the mirror state, whose covariance matrix ${\bm M}$ is associated with $\varphi_m=\pi$ (cfr. Sec.~\ref{Wigner}).

\renewcommand{\theequation}{B-\arabic{equation}}
\setcounter{equation}{0}  
\section*{APPENDIX B}  

We now give the detailed expressions of the squeezing parameter and effective mean  photon number of the mirror state. As ${\bm v}^M_{{\cal S}}$ is diagonal, it is straightforward to prove that $e^{4s_M}=\omega_mE_{\uparrow}/E_{\downarrow}$ with
\begin{equation}
\label{formulasq}
\begin{aligned}
&E_{\uparrow}=G^4\omega_m\Gamma\kappa\Delta+2N\kappa(\gamma+\kappa)\delta^2[\omega^4_m-2\omega^2_m(\gamma\kappa+\Delta^2)\\
&+z^2\delta^2]+G^2\{-2N\omega^3_m\kappa(\gamma+\kappa)\Delta+2\omega^2_m\gamma\kappa\delta^2-2\kappa^2z^2\delta^2\\
&+N\omega_m\Delta(\gamma+\kappa)[2\kappa\delta^2+\Gamma(\Delta^2-2\kappa^2)]\},\\
&E_{\downarrow}=(-G^2\Delta+\omega_m\delta^2)\{G^2[-2\omega^2_m\kappa^2+N\omega_m\Gamma(\gamma+\kappa)\Delta\\
&+2\gamma\kappa\delta^2]\}
+2N\kappa(\gamma+\kappa)[\omega^4_m-2\omega^2_m(\gamma\kappa+\Delta^2)+z^2\delta^2]
\end{aligned}
\end{equation}
with $N=2\nbar_M+1,\,\Gamma=\gamma_m+2\kappa,\,\gamma=\kappa-\Gamma$, $z=\sqrt{\gamma^2+\Delta^2}$ and $\delta=\sqrt{\kappa^2+\Delta^2}$. On the other hand, by solving $(\nbar_M+1/2)e^{2s_M}={\bm M}_{11}$ using the above expression for $s_M$ it is easy to obtain the dependence of the mean phonon number on the detuning. We get $\nbar_M=-1/2+({A_\uparrow}/{A_\downarrow})\sqrt{\omega_m{E_{\downarrow}}/{{E_\uparrow}}}$ with
\begin{equation}
\begin{aligned}
&A_{\uparrow}=-G^4\omega_m\Gamma\kappa\Delta-2N\kappa\delta^2(\gamma+\kappa)[\omega^4_m-2\omega^2_m(\gamma\kappa+\Delta^2)\\
&+z^2\delta^2]+G^2\{2N\omega^3_m\kappa\Delta(\gamma+\kappa)+2\kappa\delta^2(\kappa{z}^2-\omega^2_m\gamma)\\
&-N\omega_m\Delta(\gamma+\kappa)[(2\gamma-\Gamma)\kappa^2+(2\kappa+\Gamma)\Delta^2]\},\\
&A_{\downarrow}=2(-G^2\Delta+\omega_m\delta^2)[-2\omega^4_m\kappa(\gamma+\kappa)+G^2\omega_m\Gamma^2\Delta\\
&+4\omega^2_m\kappa(\gamma+\kappa)(\gamma\kappa+\Delta^2)-2\kappa(\gamma+\kappa)z^2\delta^2].
\end{aligned}
\end{equation}
These equations are operatively very convenient. Their explicit {\it plug-and-{play}} nature allows to quantify the expected degree of squeezing and the quantitative thermal character of the mirror state per assigned experimental configuration.


\begin{thebibliography} {99}
\bibitem{progressi} M. D.\ LaHaye {\it et al.}, Science {\bf 304}, 74 (2004), S.\ Mancini {\it et al.}, \prl {\bf 88}, 120401 (2002); A. D.\ Armour {\it et al.}, {\it ibid} 148301 (2002); W.\ Marshall {\it et al.}, {\it ibid} {\bf 91}, 130401 (2003).
\bibitem{david} D. Vitali, S. Gigan, A. Ferreira, H. R. B\"ohm, P. Tombesi, A. Guerreiro, V. Vedral, A. Zeilinger, and M. Aspelmeyer, {Phys. Rev. Lett.} {\bf 98}, 030405 (2007).
\bibitem{nano}  A. Naik, O. Buu, M. D. LaHaye, A. D. Armour, A. A. Clerk, M. P. Blencowe, and K. C. Schwab, Nature (London), {\bf 443}, 193 (2006).
\bibitem{altri} F. Marquardt, J. P. Chen, A. A. Clerk, and S. M. Girvin, Phys. Rev. Lett. {\bf 99}, 093902 (2007); I. Wilson-Rae, N. Nooshi, W. Zwerger, and T. J. Kippenberg, Phys. Rev. Lett. {\bf 99}, 093901 (2007); F. Xue, Y. D. Wang, Y.xi Liu, and F. Nori, Phys. Rev. B {\bf 76}, 205302 (2007).
\bibitem{ionjp} M. Paternostro, S. Gigan, M. S. Kim, F. Blaser, H. B\"ohm, and M. Aspelmeyer, {New J. Phys.} {\bf 8}, 107 (2006).
\bibitem{cool} S. Gigan, H. R. B\"ohm, M. Paternostro, F. Blaser, G. Langer, B. Hertzberg, K. Schwab, D. B\"auerle, M. Aspelmeyer, and A. Zeilinger, {Nature (London)}, {\bf 444}, 67 (2006); O. Arcizet, P.-F. Cohadon, T. Briant, M. Pinard, and A. Heidmann, {ibidem}, 71 (2006); D. Kleckner and D. Bouwmeester, {ibidem}, 75 (2006); A. Schliesser, P. Del'Haye, N. Nooshi, K. J. Vahala, and T. J. Kippenberg, {Phys. Rev. Lett.} {\bf 97}, 243905 (2006); A. Schliesser, R. Rivi\'ere, G. Anetsberger, O. Arcizet, and T. J. Kippenberg, Nature Phys. {\bf 4}, 415 (2008); T. Corbitt, Y. Chen, E. Innerhofer, H. M\"uller-Ebhardt, D. Ottaway, H. Rehbein, D. Sigg, S. Whitcomb, C. Wipf, and N. Mavalvala, Phys. Rev. Lett. {\bf 98}, 150802 (2007); J. D. Thompson, B. M. Zwickl, A. M. Jayich, F. Marquardt, S. M. Girvin, and J. G. E. Harris, Nature {\bf 452}, 72 (2008); S. Gr\"oblacher, S. Gigan, H. R. B\"ohm, A. Zeilinger, and M. Aspelmeyer, Europhys. Lett. {\bf 81}, 54003 (2008); A. Schliesser, O. Arcizet, R. Rivi\'ere, G. Anetsberger, and T. J. Kippenberg, Nature Phys. {\bf 5}, 509 (2009); S. Gr\"oblacher, J. B. Hertzberg, M. R. Vanner, G. D. Cole, S. Gigan, K. C. Schwab, and M. Aspelmeyer, {\it ibid.} {\bf 5}, 485 (2009); Y.-S. Park and H. Wang, {\it ibid.} {\bf 5}, 489 (2009); A. D. O'Connell, {\it et al.}, Nature (London) {\bf 464}, 697 (2010); J. D. Teufel, T. Donner, Dale Li, J. H. Harlow, M. S. Allman, K. Cicak, A. J. Sirois, J. D. Whittaker, K. W. Lehnert, R. W. Simmonds, {\it ibid.} {\bf 475}, 359 (2011); J. Chan, T. P. Mayer Alegre, A. H. Safavi-Naeini, J. T. Hill, A. Krause, S. Groeblacher, M. Aspelmeyer, and O. Painter, {\it ibid.} {\bf 478}, 89 (2011); J. Chan, {\it et al.}, Appl. Phys. Lett. {\bf 101}, 081115 (2012).
\bibitem{loro} S. Gr\"oblacher, K. Hammerer, M. R. Vanner, and M. Aspelmeyer, Nature {\bf 460}, 724 (2009).
\bibitem{Safavi} A. Safavi-Naeini, {\it et al.}, Phys. Rev. Lett. {\bf 108}, 033602 (2012).
\bibitem{boseferreira} S. Bose, K. Jacobs, and P. L. Knight, \pra {\bf 56}, 4175 (1997); S. Mancini, V. I. Man'ko, and P. Tombesi, Phys. Rev. A {\bf 55}, 3042 (1997); A. Ferreira, A. Guerreiro, and V. Vedral, Phys. Rev. Lett. {\bf 96}, 060407 (2006). 
\bibitem{3Bpaper} M. Paternostro, D. Vitali, S. Gigan, M. S. Kim, C. Brukner, J. Eisert, and M. Aspelmeyer, Phys. Rev. Lett. {\bf 99}, 250401 (2007).
\bibitem{law} C. K. Law, Phys. Rev. A {\bf 51}, 2547 (1995). 
\bibitem{eisertplenio} J. Eisert and M. B. Plenio, Int. J. Quant. Inf. {\bf 1}, 479 (2003).
\bibitem{benguria} R. Benguria and M. Kac, Phys. Rev. Lett. {\bf 46}, 1 (1981).
\bibitem{commentocomparison} We remark that finding the state of the system resulting from this open dynamics is a formidable task. The possibility of using finite-dimensional covariance matrices instead illustrates the power of the covariance matrix formalism for Gaussian states.
\bibitem{io} M. Paternostro, J. Phys. B {\bf 41}, 155503 (2008).
\bibitem{inpout} M. J. Collett and C. W. Gardiner, \pra{\bf 30}, 1386 (1984).
\bibitem{vitaliout} C. Genes, A. Mari, P. Tombesi, and D. Vitali, Phys. Rev. A {\bf 78}, 032316 (2008). 
\bibitem{PulsedTheo} M. R. Vanner, I. Pikovski, G. D. Cole, M. S. Kim, C. Brukner, K. Hammerer, G. J. Milburn, M. Aspelmeyer, Proc. Nat. Acad. Sci. USA {\bf 108}, 16182 (2011). 
\bibitem{PulsedExp} M. R. Vanner, J. Hofer, G. D. Cole, and M. Aspelmeyer, arXiv:1211.7036 (2012).
\bibitem{eisert} J. Eisert, S. Scheel, and M. Plenio, {\prl} {\bf 89}, 137903 (2002).
\bibitem{footnote} {In order to apply the formulae for updated covariance matrices, we need to adapt ${\bf V}$ to the convention where the quadratures of one mode in the vacuum state all have variances equal to $1$~\cite{eisert}. This accounts to take twice the value of each element in Eq.~(\ref{tipica}).}
\bibitem{giedke} G. Giedke, B. Kraus, M. Lewenstein, and J. I. Cirac, \pra {bf 64}, 052303 (2001).
\bibitem{barnettradmore} S. M. Barnett and P. M. Radmore, {\it Methods in Theoretical QUantum Optics} (Oxford University Press, New York, 1997).
\bibitem{ent} Logarithmic negativity is an entanglement monotone and can be calculated using the symplectic spectrum of the matrix ${\bf V}^{P}={\bf P}\,{\bf V}{\bf P}$, which is the covariance matrix associated with the partially transposed density matrix of the system. Here, ${\bf P}=\openone\oplus{\bm \sigma}_z$ (with $\openone$ the $2\times{2}$ identity matrix, ${\bm \sigma}_r$ the $r$-Pauli matrix and $r=x,y,z$)~\cite{david,3Bpaper}.
\end{thebibliography}
\end{document}